\documentclass[journal,transmag]{IEEEtran}

\usepackage{cite}
\usepackage{graphicx}
\usepackage{amssymb,amsmath}

\begin{document}

\title{Skyrmion vs. antiskyrmion Hall angles}


\author{
	\IEEEauthorblockN{Bom Soo Kim}
	\IEEEauthorblockA{Department of Mathematics and Physics, University of Wisconsin-Parkside,  Kenosha WI 53141, USA, \\ kimb@uwp.edu}
}

\IEEEtitleabstractindextext{%
\begin{abstract}
	We observe significant, 6.5\% - 12\%, differences between skyrmion and antiskyrmion Hall angles in existing experimental data. To account for them, the Thiele equation is generalized with the transverse velocity component in the collective coordinate of magnetization vector. The corresponding generalized Hall angle is formulated, and its consequences are compared to the data. A plausible physical origin is provided for explaining the Hall angle differences. 
\end{abstract}

}

\maketitle

\pagestyle{empty}
\thispagestyle{empty}

\IEEEpeerreviewmaketitle

\section{Introduction}

\IEEEPARstart{S}{kyrmions} are particle-like extended spin textures that have attracted much attention due to small size, topological protection and low operational costs. One of their signature properties is the skyrmion Hall effect, that pushes skyrmion and antiskyrmion toward opposite transverse directions. This Hall effect can be conveniently described by the Hall angle $\tan \theta_H = v_\perp / v_\parallel $ where $v_\perp$ and $ v_\parallel $ are transverse and longitudinal velocities with respect to an applied current. There exist small number of systematic experimental Hall angle data that contain both skyrmion and antiskyrmion \cite{Jiang2017}\cite{VanishingSkyrmionHall}. 

Attentive observations reveal unexpected disparity between the skyrmion Hall angle and the antiskyrmion Hall angle with the same experimental environments. A clean data set is provided in  \cite{VanishingSkyrmionHall} for the N\'eel-type half skyrmions (with the topological skyrmion charge $Q=2\pi$) and antiskyrmions ($Q=-2\pi$) in the ferrimagnetic GdFeCo/Pt film by utilizing the spin orbit torque (SOT) technique. At temperature $T=343 \, K$, the Hall angles are given as 
\begin{align} \label{DataAt343K}
	\theta_{SkH} = - 35^o \;,  \qquad  \theta_{ASkH} = 31^o \;,
\end{align}
with their $\% \, \text{difference}$ as   
\begin{align}
	\% \, \text{difference} = 12 \% \;.
\end{align}
This is a strikingly large difference with the naive expectation that the two Hall angles are supposed to be the same. 

Also, in one of the earliest measurements for the skyrmion Hall effect \cite{Jiang2017}, the Hall angles for N\'eel-type (anti)skyrmion for Ta/CoFeB/TaO$_x$ were measured using the spin Hall effect with opposite applied currents $\pm J_\parallel $ and (unfortunately) different magnetic fields $B$. It was required to use strong current pulses due to the material's strong pinning effects. Among numerous data points presented in \cite{Jiang2017}, we use only the saturated Hall angle data that are further interpolated to get the data points for $B=\pm 5.0 \, Oe $ \cite{Kim:2020piv}. The resulting Hall angles are $\theta_{SkH} = -31.6^o $ and $\theta_{ASkH} = 29.3^o $ with their $\% \, \text{difference} = 7.6 \%$ for $+J_\parallel$. For the opposite current $-J_\parallel$, similar analysis leads to $\% \, \text{difference} = 5.5 \%$. Putting them together,  the conservative estimate for  \% difference between skyrmion and antiskyrmion Hall angles is $ 6.5\%$. 

Thus we see that skyrmion Hall angles are consistently much larger (by 6.5\% -12\%) than those of the antiskyrmion.  

\section{Generalized Hall angle} 

In a seminal paper \cite{Thiele}, Thiele formulated the equation for describing the steady state motion of a skyrmion center, which is parametrized by the collective coordinate $X_i = v_i t$ with a longitudinal velocity $v_i$ with a vector index $i$. The normalized magnetization vector is  
\begin{align}
	n_i = M_i ( x_j - X_j)/M_s \;, 
\end{align} 	
where $M_s = |\vec M| $ is a saturation magnetization and $x_i$ the field position. One can check $M_i \partial_t M_i =  0$ and $ M_i \partial_j M_i = 0$ due to the normalization $\vec n^2 =1$. The repeated indices are summed over. The resulting Thiele equation is 
\begin{align} \label{ThieleEQ}
	\mathcal G_{ij} v_j + \alpha \mathcal D_{ij} v_j + F_i = 0 \;,
\end{align} 
here we consider 2 dimensions ($i,j,k= x, y$).   
$\mathcal G_{ij} =\epsilon_{ij} Q $ is related to the Magnus force with the skyrmion charge $Q= \epsilon_{ijk} \int d^2 x n_i  (\partial_x n_j) (\partial_y n_k) $. $\mathcal D_{ij} = \delta_{ij} D $ is dissipative force with $D= - \int d^2 x (\partial_x n_k) (\partial_x n_k) $ that is symmetric and has the same values for diagonal components for the fundamental N\'eel or Bloch skyrmions. $ F_i = -\gamma_0 \int d^2 x (\partial_i n_j) H_j$ is external force that can include forces due to various spin torques. The integrals are over the unit Skyrmion volume. Internal forces due to anisotropy and exchange energies, internal demagnetizing fields, magnetostriction do not contribute. 

For $\vec F = F \hat x $ without loss of generality, we get the Hall angle from \eqref{ThieleEQ} (for $i=y $) as   
\begin{align} \label{SkHangle}
	\tan \theta_{H} = \frac{v_y}{v_x} =\frac{Q }{\alpha D} \;. 
\end{align}	
Note that this Hall angle cannot explain the disparity described above as it gives the same value for skyrmion and antiskyrmion with opposite signs.

To accommodate the disparity between skyrmion and antiskyrmion Hall angles, we generalize the collective skyrmion coordinate $X_i$ as \cite{Kim:2020piv}\cite{KimBook}
\begin{align} \label{SkyrmionCM}
	X_i = v_i t + R \epsilon_{ij} v_j t \;.
\end{align} 
Without altering the constraints, $M_i \partial_t M_i =  0$, and $ M_i \partial_j M_i = 0$, the Thiele equation \eqref{ThieleEQ} is generalized to 
\begin{align} \label{ThieleEQGeneral}
	\mathcal G_{ij} (v_j + R \epsilon_{jk} v_k) + \alpha \mathcal D_{ij} (v_j + R \epsilon_{jk} v_k) + F_i = 0 \;,
\end{align} 
where the new terms come from $\partial_t M_i = (v_j + R \epsilon_{jk} v_k) \partial_j M_i $. 

\newpage 

\begin{figure}[t!]
	\begin{center}
		\includegraphics[width=0.2\textwidth]{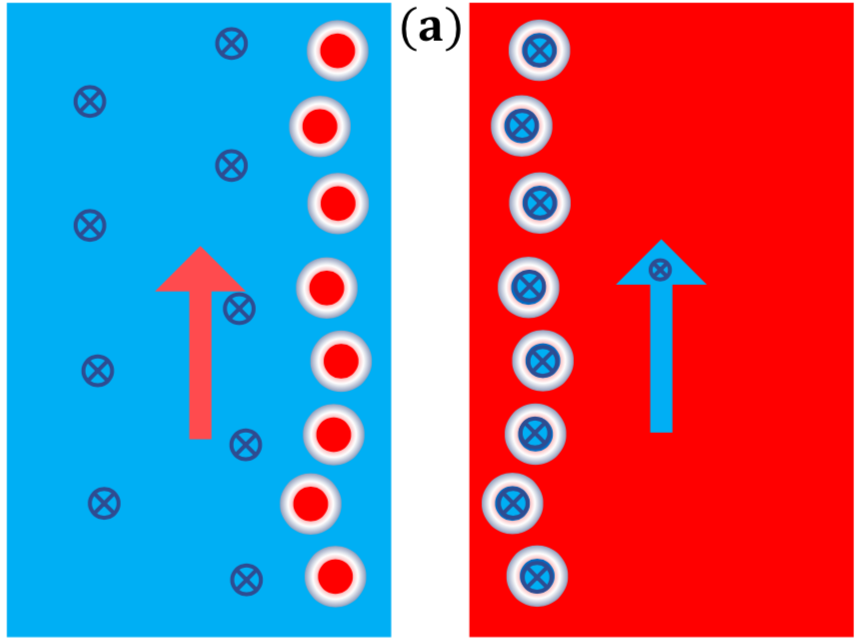} ~~
		\includegraphics[width=0.2\textwidth]{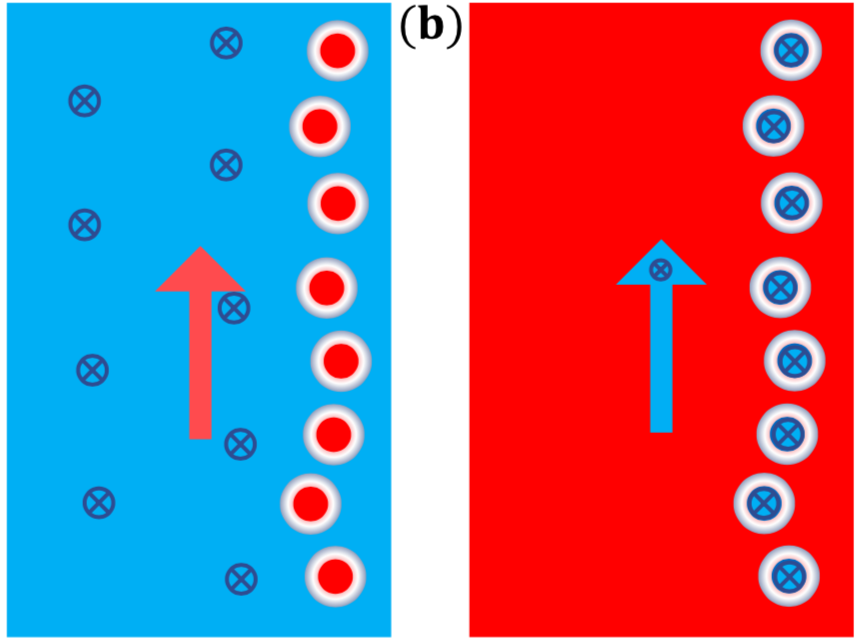} 
		\caption{(a) Skyrmion Hall effect: skyrmions (red dots) in down-spin ferromagnet (blue background) move toward right, while antiskyrmions (blue dots) in up-spin ferromagnet toward left. (b) $DR$: both skyrmions and antiskyrmions move toward right. Thick arrow is the direction of a driving force. These two effects are added to the skyrmions, while subtracted to antiskyrmions.     \vspace{-0.2in} 
		}
		\label{fig:HallDrag}
	\end{center}
\end{figure}

There are two new terms. (1) $\mathcal G_{ij} R \epsilon_{jk} v_k = - Q R v_i $ is parallel to the applied force $F_i$. It is proportional to the skyrmion charge and thus either an accelerating or decelerating force. (2) $\mathcal D_{ij} R \epsilon_{jk} v_k = D R \epsilon_{ik} v_k $ is an additional transverse contribution added to $Q$. This turns out to be quadratic to the magnetization vector $\vec n$. Thus it is independent of the skyrmion charge and universal transverse force pushing both skyrmion and antiskyrmion toward the same transverse direction. 

The generalized Hall angle is 
\begin{align} \label{SkHangleGeneral}
	\tan \theta_{H} = \frac{Q + \alpha D R}{\alpha D - QR } \;. 
\end{align}	
One can check it reduces to \eqref{SkHangle} for $R=0$. This include the new universal transverse contribution $\alpha D R$ and the longitudinal one $ - QR $ as well. 

\section{Compared to Experimental Data} 

\begin{figure}[b!]
	\begin{center}
		\includegraphics[width=0.35\textwidth]{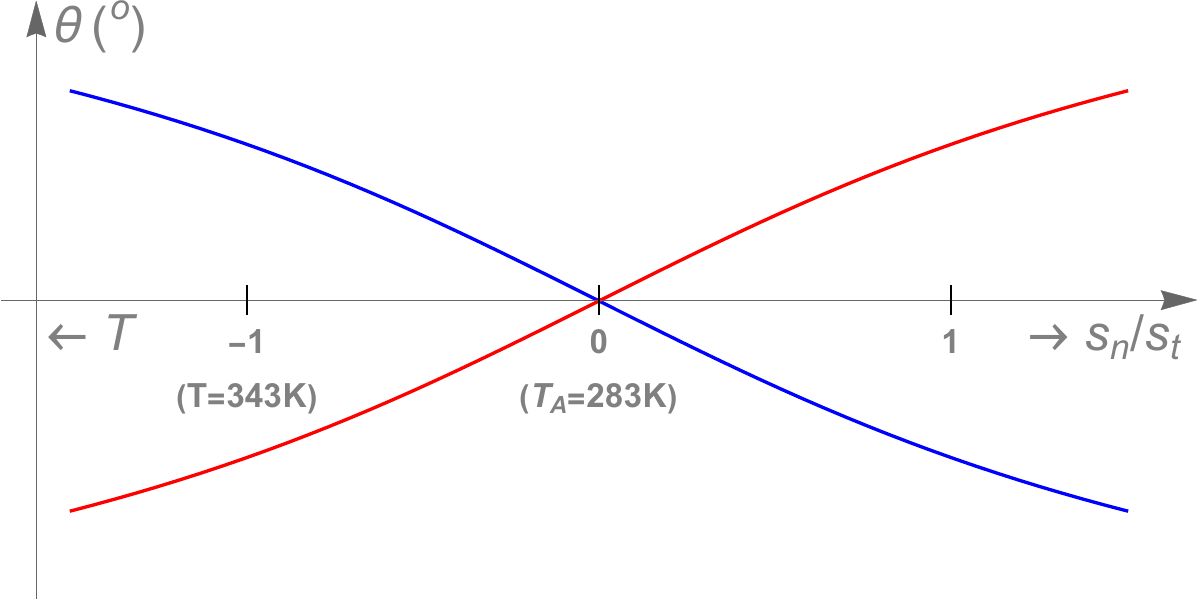} 
		\caption{$\theta_{H}$ of \eqref{SkHangleSnet} as a function of $s_n/s_t $, which is a decreasing function of $T$. This captures essential features of experimental data in Fig. 4 \cite{VanishingSkyrmionHall}. 
			\vspace{-0.2in} 
		}
		\label{fig:HallAngleSpinCompared10A}
	\end{center}
\end{figure}

We estimate $R$ in \eqref{SkHangleGeneral} from experimental data given in \eqref{DataAt343K} \cite{VanishingSkyrmionHall}. Applying \eqref{SkHangle} twice for skyrmion and antiskyrmion Hall angles, we obtain $R= -0.035$ and $\alpha D = -9.68$. Note $\alpha D R > 0$ and added to $Q>0$ for skyrmions in \eqref{SkHangleGeneral}. See Fig. \ref{fig:HallDrag}. The strength of the new transverse force $\alpha \mathcal D_{ij}  R \epsilon_{jk} v_k$ compared to the skyrmion Hall effect $\mathcal G_{ij} v_j$ is given by \cite{Kim:2020piv}\cite{KimBook}
\begin{align} \label{HVEstimate}
	\frac{\alpha D R}{Q} = 5.4\% \;.
\end{align}	
It is a surprisingly large contribution. Similar analysis with the data mentioned above \cite{Jiang2017} gives $\alpha D R / Q = 2.9\%$.  

After seeing the significant contributions of $\alpha D R$, we explorer consequences of the existence of the transverse velocity component $R$ near the angular momentum compensation point $T_A$ of ferrimagnet GdFeCo/Pt considered again in \cite{VanishingSkyrmionHall}. Ferrimagnet compound is composed with two subnetworks of magnetization (anti-ferromagnetically coupled) whose magnetic moments $M_a, a=1,2$ and spin densities $s_a$ depend on temperature $T$. See \cite{FerrimagnetFastMovongDomainWall} for a general theory. At $T=T_A$, two spin densities cancel each other. Moreover the ratio of the net spin density and total spin density $ s_n/s_t = (s_1 - s_2)/(s_1 + s_2)$ is monotonically decreasing function of temperature. 

\cite{VanishingSkyrmionHall} modeled the Magnus force as $ \vec F_g = s_n Q (\hat z \times \vec v)$ and the viscose force as $ \vec F_d = s_t \alpha D \vec v$. Thus, the spin (and temperature) dependence can be incorporated by $Q \to s_n Q$ and $\alpha D \to s_t \alpha D $. Thus, the Hall angle \eqref{SkHangleGeneral} turns into 
\begin{align} \label{SkHangleSnet}
	\tan \theta_{H}= \frac{s_n Q+ s_t \alpha D R}{s_t \alpha D - s_n QR } \;. 
\end{align}	
Repeating analysis, $R= -0.035$ and $(s_t/s_n) \alpha D = -9.68$. The Hall angles are plotted in Fig. \ref{fig:HallAngleSpinCompared10A} as a function of $s_n/s_t$, with a specification $s_t/s_n =-1 $ happens at $T=343 \, K$. 

\begin{figure}[t!]
	\begin{center}
		\includegraphics[width=0.34\textwidth]{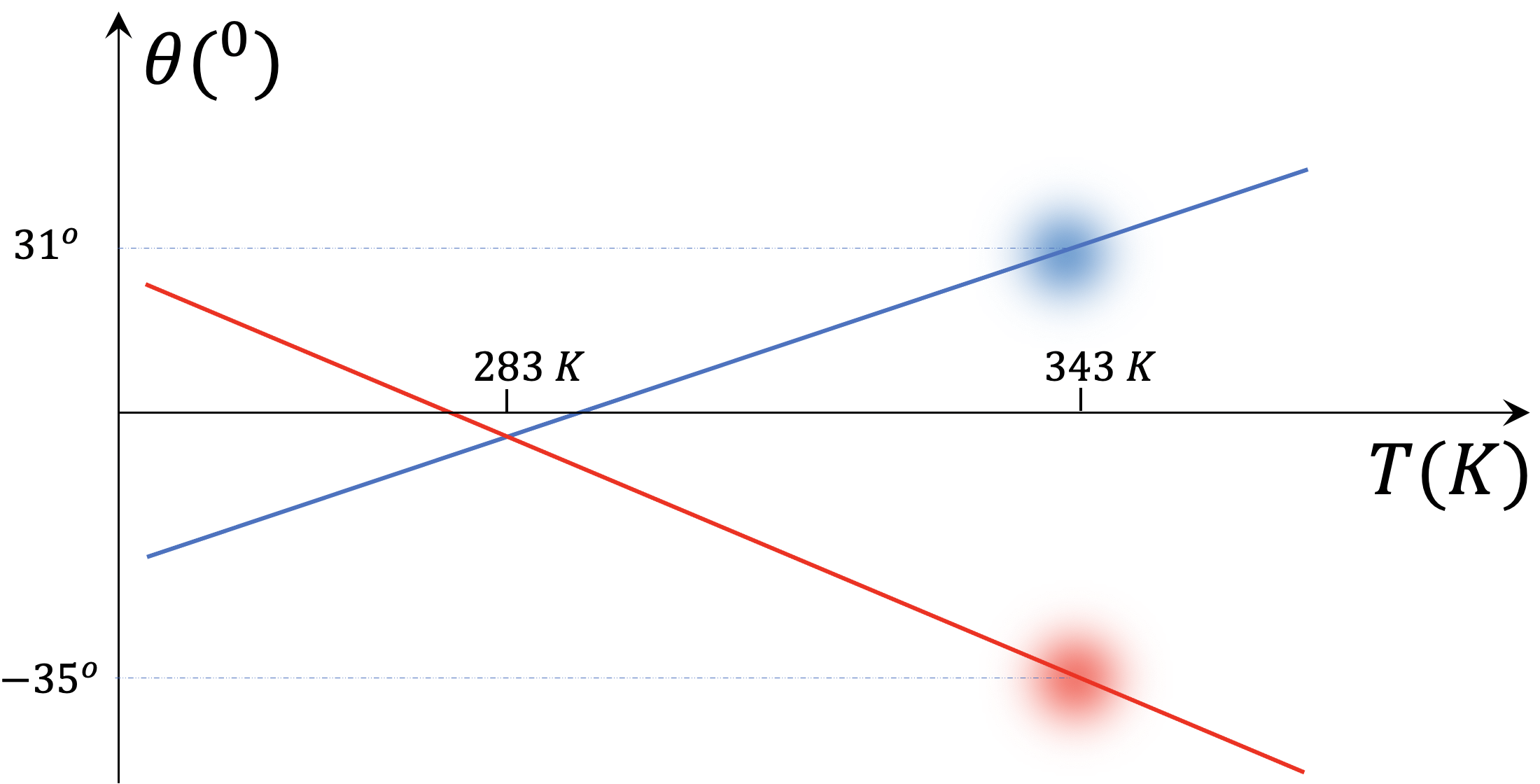} 
		\caption{A triangle formed by skyrmion (red color) and antiskyrmion (blue) Hall angles near the compensation point. Compare this with Fig. 3 in \cite{VanishingSkyrmionHall}.      \vspace{-0.32in} 
		}
		\label{fig:AMCTTriangle2}
	\end{center}
\end{figure}

We observe the skyrmion Hall angle vanishes when $ s_n/s_t > 0 $, while that of antiskyrmion does when $ s_n/s_t < 0 $. Thus skyrmion Hall angle vanishes at a smaller temperature than the antiskyrmion. The compensation temperature $ s_n (T_A) = 0$ lies in between them, and the Hall angles are the same $\tan \theta_{H}= R < 0$. Thus there is a triangle structure formed by the Hall angles near $T_A$, illustrated in Fig. \ref{fig:AMCTTriangle2} as a function of $T$. 

\section{Conclusion} 

By introducing the transverse velocity component, we account for the surprising disparity between the skyrmion and antiskyrmion Hall angles. The corresponding universal transverse force estimates to be $2.9\% - 5.4\% $ of skyrmion Hall effect. This can play crucial roles for fast moving skyrmions in, say, the next generation skyrmion racetrack devices.    

Is there a candidate responsible for this universal transverse force? Yes! The so-called Hall viscosity was introduced in skyrmion physics in \cite{Kim:2015qsa} through utilizing QFT Ward identity. It is transverse to skyrmion motion and shown to be independent of its charge using Kubo formula \cite{Kim:2020piv}. Various ways to verify Hall viscosity in skyrmion motions were proposed in \cite{Kim:2019vxt}\cite{KimBook}. 

We advertise to perform systematic precision experiments for skyrmion and antiskyrmion together to verify the universal transverse force and Hall viscosity.

\end{document}